\documentclass[runningheads]{llncs}
\usepackage[T1]{fontenc}
\usepackage{graphicx}
\usepackage{amsmath}
\usepackage{amsfonts}
\usepackage{amssymb}
\usepackage{multirow}
\usepackage{physics}

\usepackage[
backend=biber,
style=numeric,
sorting=none
]{biblatex}
\usepackage{xcolor}
\graphicspath{{./Figs/}}
\begin{document}
\title{Coarsening and Bifurcations in Wide-Range Two-Dimensional Totalistic Cellular Automata}
\titlerunning{Wide-Range Two-Dimensional Totalistic CA}
\author{Franco Bagnoli\inst{1,2\dagger}\orcidID{0000-0002-6293-0305} \and
Luca Mencarelli\inst{1}
}
\authorrunning{F. Bagnoli and L. Mencarelli}

\institute{Department of Physics and Astronomy and CSDC, University of Florence,  via G. Sansone 1, 50019 Sesto Fiorentino (Italy) \email{franco.bagnoli@unifi.it}, 
\email{luca.mencarelli3@edu.unifi.it}
\and INFN, Sect. Florence (Italy) 
}
\maketitle             

\let\at@
\catcode`@=\active
\def@#1{\ifmmode\boldsymbol{#1}\else\at#1\fi}

\let\quot"
\catcode`"=\active
\def"#1"{``#1''}

\newcommand{\eq}[2][]{%
    \ifthenelse{\equal{#1}{}}{%
        \begin{equation*}
            #2%
        \end{equation*}%
    }{%
        \begin{equation}\label{eq:#1}%
            #2%
        \end{equation}%
    }%
}
\newcommand{\meq}[2][]{%
    \ifthenelse{\equal{#1}{}}{%
        \begin{equation*}%
            \begin{split}%
                #2%
            \end{split}%
        \end{equation*}%
    }{%
        \begin{equation}\label{eq:#1}%
            \begin{split}%
                #2%
            \end{split}%
        \end{equation}%
    }%
}

\begin{abstract}
We investigate Boolean, totalistic cellular automata with a majority or frustrated majority vote rule, and an interaction range of variable span. These two models show a behavior which differs from the mean-field one. The majority vote model is characterized by the presence of absorbing states, and there is a related bifurcation according to the initial density, in agreement with the mean-field approximation. For initial density equal to $0.5$, however, the dynamics is dominated by a coarsening process, which stops when clusters with a definite curvature radius are established. For the frustrated majority vote model, the mean-field approximation gives chaotic oscillations or a limit cycle. Instead, we observe active patterns, with stable density. Above a certain critical value for the interacting radius there is a bifurcation of the asymptotic density as a function of the initial one. 
\end{abstract}


\section{Introduction}

The study of totalistic, deterministic cellular automata (CA) is a rich, interdisciplinary field spanning statistical physics and discrete dynamical systems~\cite{Wolfram1983}, with many applications~\cite{ACRI}. 

At its core, a totalistic rule implies that the future state of the cell depends symmetrically on the present one of cells belonging to its neighborhood. 

A particular case is the majority vote rule, which dictates that a cell takes on the state held by the majority of its neighbors~\cite{Myczkowski1989}. This rule is analogous to the voter model~\cite{holley1975ergodic}, which, as we shall argue in the following, can be considered as a stochastic version of the majority rule. It can also be considered as the zero-temperature limit of the Ising Model~\cite{Bray1994}, and has a natural interpretation in terms of opinion dynamics~\cite{Castellano2009}. 

In two dimensions, the dynamics of the  nearest-neighbors majority model, like that of the Ising model, is determined by the Cahn-Allen theory of coarsening due to the curvature of the front and a kind of  surface tensions~\cite{Bray1994}. However, although one can apply a scaling approach to study systems with larger interacting radius, a study of majority rule with long-range interactions is missing. 

In any majority model, there are at least two absorbing states, constituted by a homogeneous configuration of sites. We shall show that there are also other stable configurations, formed by clusters of sites whose boundaries present a curvature related to the interacting radius, in a non-trivial way. These configurations are not forecast by a mean-field analysis, and are unstable under stochastic perturbation, even when these obeys the same symmetry law of the original rule, like in the voter model.

Another interesting phenomenon happens if we remove the presence of absorbing states, still using a deterministic approach. In this case the mean-field approximation gives a chaotic dynamics, but instead what is observed in numerical simulations is a bifurcation in the asymptotic density with respect to the initial one, for a large enough  interaction range. 

The outline of this work is the following: in Section~\ref{sec:model} we present the mathematical definition of the models and the related mean-field approximations. In Section~\ref{sec:majority} we report the results of numerical simulation of the majority model, with some analytical approximation to the problem of determining the curvature radius. In Section~\ref{sec:frustrated} we show the numerical results for the frustrated majority model. Conclusions are drawn in the last section. 

\section{Definitions}\label{sec:model}

A cellular automaton (CA) is defined by a set of $N$ cells, indexed by a discrete index $i$, whose state $s_i$  can be one of a number of possible integer numbers. In this work we consider Boolean CA, for which $s_i\in\{0,1\}$.
We denote the state of all cells in the set at a given time by $@s\equiv @s(t)$. 

Each cell evolves its state according to the state of other cells (possibly including the cell itself) at the previous time step. The geometry of the interactions is given by an adjacency matrix $a_{ij}$, which takes value one if cell $i$ depend on cell $j$ and zero otherwise. 

We consider here two-dimensional square lattices of cells, of dimensions $N=X\times Y$, so that, given a cell $i$ of coordinates $(x_i,y_i)$, we have 
\eq{
    i = y_i X + x_i; \qquad x_i = i \bmod X; \qquad y_i = \lfloor i / X\rfloor.
}

Let us define the distance between two sites,  
\eq{
    d_{ij} = \sqrt{(x_i-x_j)^2 + (y_i-y_j)^2}.
}

We introduce the interacting radius $R$ such that 
\eq{
    a_{ij} = \begin{cases} 
     1 & \text{if $d_{ij} \le R$,}\\
     0 & \text{otherwise.}
     \end{cases}
}
The set of cells that are interacting with a given one is called its neighborhood $@v_i$
\eq{   
    @v_i = \{s_j \;|\; a_{ij}=1\}.
}
We consider periodic boundary conditions, so all operations on the $x$ and $y$ coordinates are, respectively, modulo $X$ and $Y$. 

We introduce  the totalistic state $\tau_i$ of the neighborhood of cell $i$ as 
\eq{
    \tau_i = \sum_j a_{ij} s_j.
}
Notice that, by symmetry, the number of cells in the neighborhood, 
\eq{
    K = k_i = \sum_j a_{ij},
} 
is always odd (since the lattice is regular, this number is the same for all cells). 

\begin{figure}[t]
    \centering
    \includegraphics[width=0.5\linewidth]{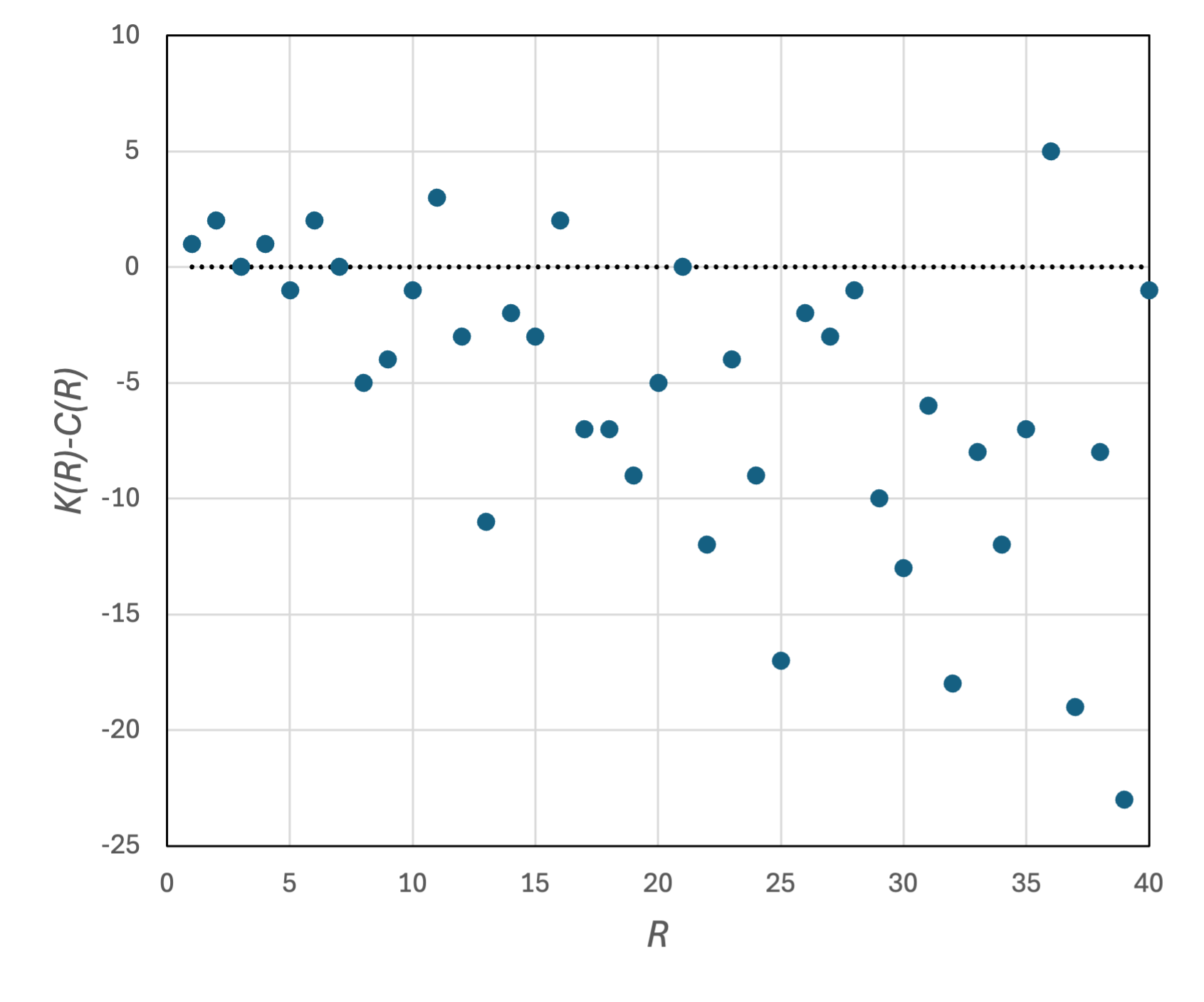}
    \caption{
    The difference $K(R)-C(R)$ between the number of point $K(R)$ within distance $R$ from the origin in a regular two-dimensional grid of unitary spacing, and the approximation $C(R)=\lfloor \pi R^2\rfloor$, for $0\le R\le 40$. }\label{fig:K}
\end{figure}

The relationship between the interacting radius $R$ and $K(R)$ is not trivial, it is the so-called Gauss circle problem~\cite{wiki:GaussCircleProblem}, and, while $K(R)$ is clearly approximated by $C(R)=\lfloor \pi R^2\rfloor$, this approximation shows fluctuations which in average increase with $R$, as shown in Figure~\ref{fig:K}.

The state of each cell evolves according to a function $S(v)$ of its neighborhood,
\eq{
    s'_i = S(@v_i),
}
this rule can be applied simultaneously to all cells, or serially. For serial simulations, cells are updated in a random order, so the number of individual updates per time step is the same in both cases. 

In the case of totalistic rules, the function actually depends on the totalistic state of the neighborhood $S(@v_i)=S(\tau_i)$.

The observable we consider is the average value of the state of cells (the ``density''), 
\eq{
    \rho(@s) = \frac{1}{N} \sum_i s_i.
}

The mean-field approximation of the evolution is obtained by neglecting all correlations (i.e., shuffling the configuration or choosing random neighbors -- which corresponds to shuffling the rows of $a_{ij}$). 
If we disregard correlations, the probability of finding a random cell in state 1 corresponds to the density $\rho$. 

The mean-field evolution equation for the density is
\eq[mf]{
    \rho' = \sum_{v=0}^K \binom{K}{v} S(v) \rho^v (1-\rho)^{K-v}.
}

The stationary points of this discrete-time difference equation are given by $\rho^*=\rho'=\rho$, and they are stable if
\eq{
    \left|\dv{\rho'}{\rho}\right|_{\rho=\rho^*} < 1.
}

The majority rule $M(v)$ is defined as 
\eq[majority]{
    M(v) = \begin{cases}
        1 & \text{if $v > K/2$,}\\
        0 & \text{otherwise, i.e., $v < K/2$.}
        \end{cases}
}
Notice that, since $K$ is odd, the case $v=K/2$ can never happen. 
Clearly, if all cells have state 1 (0), the sequential or serial update does not change anything, so the states $\rho=0$ or $\rho=1$ are absorbing.

\begin{figure}[t]
    \centering
    \begin{tabular}{ccc}
    \includegraphics[width=0.33\linewidth]{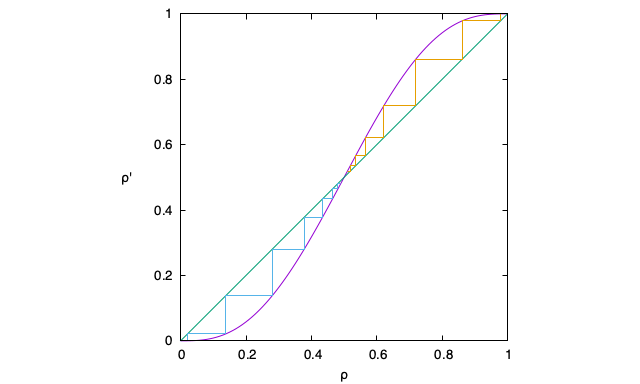} & 
    \includegraphics[width=0.33\linewidth]{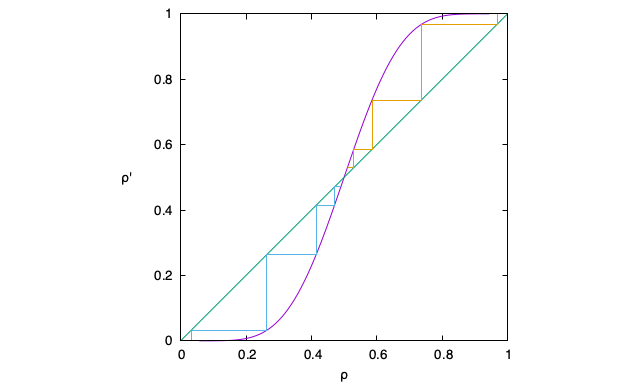} &
    \includegraphics[width=0.33\linewidth]{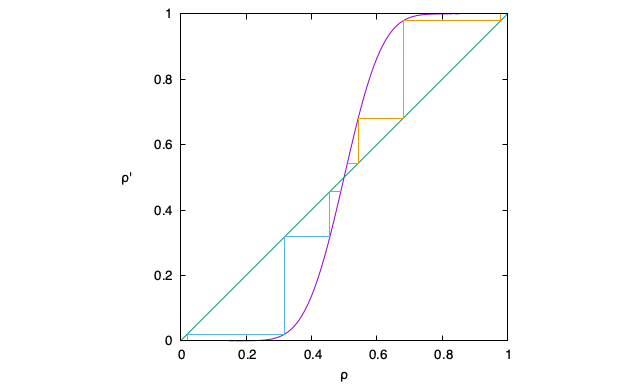} 
    \end{tabular}
    \caption{The mean-field return map of the majority model, Eq.~\eqref{eq:majority}, and a few iterations of the map. Left:  $R=1$ ($K=5$), center: $R=2$ ($K=13$), right: $R=3$ ($K=29$).}
    \label{fig:majoritymf}
\end{figure}

The behavior of the mean-field approximation is shown in Fig.~\ref{fig:majoritymf}, and it is evident that the only stable fixed states are the absorbing ones, $\rho =0$ or $\rho=1$,  while the fixed point $\rho=0.5$ is unstable. 

A related model is the voter one~\cite{holley1975ergodic,Castellano2009}, which is totalistic ``in average''. In this case each cell copies the state of a random neighbor (in the  radius-$R$ circle), so the evolution equation is
\eq{
    s'_i = V(@v) = v_j \; | \; j\; \text{random, with}\; d_{ij} \le R.
}
The mean-field equation for the voter model is 
\eq{
    \rho' = \sum_{v=0}^K \frac{v}{K} \binom{K}{v}  \rho^v (1-\rho)^{K-v} = \; \text{(performing the summation)}\; = \rho, 
}
so that in average the density does not change, disregarding fluctuations. This is consistent with the fact that the linear voter model can be mapped onto a random walk with absorbing boundaries~\cite{holley1975ergodic,Frachebourg1996,Castellano2009}, and therefore, for finite lattices in 2D, the final state of the voter model is the homogeneous all-zero or all-one configuration.

Finally, we are interested in examining what happens if we remove the absorbing state, defining a frustrated majority rule $F(v)$ as 
\eq[frustrated]{
    F(v) = \begin{cases}
        1 &\text{if $v=0$ or  $K/2 < v \le R$,}\\
        0 &\text{otherwise, i.e., $v=K$ or $1\le v<K/2$.}
    \end{cases}
}

\begin{figure}[t]
    \centering
    \begin{tabular}{ccc}
    \includegraphics[width=0.33\linewidth]{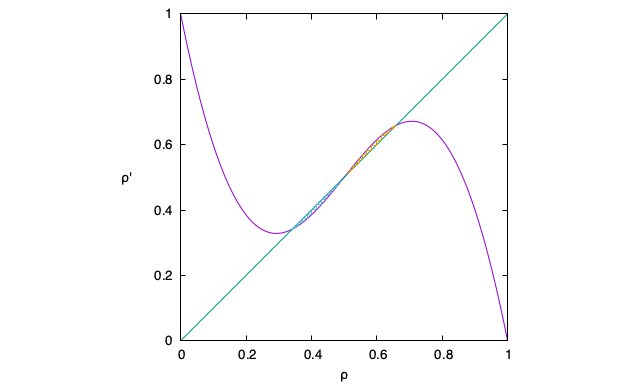} &\includegraphics[width=0.33\linewidth]{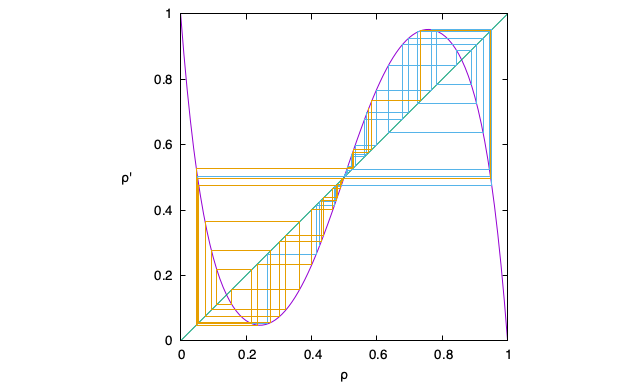} &
    \includegraphics[width=0.33\linewidth]{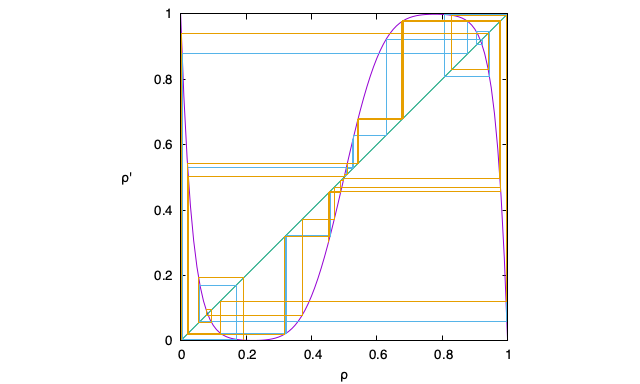} 
    \end{tabular}
    \caption{The mean-field approximation function $\rho'=F(\rho)$ for the frustrated majority model with (left) $R=1$ ($K=5$), (center) $R=2$ ($K=13$), (right) $R=3$ ($K=29$), with some iterations of the map.}
    \label{fig:frustratedmf}
\end{figure}

The mean-field map, reported in Fig.~\ref{fig:frustratedmf}, shows that the fixed point $\rho=0.5$ is always unstable, as also the limit cycle oscillating between $\rho=0$ and $\rho=1$. For $R=1$ there are two other stable points, but for larger values of $R$ the mean-field trajectories are chaotic. Numerically, it happens that for $R$ equal or larger than 5 ($K>80$), the mean-field map is so close to $\rho'=0$ or $\rho'=1$ that the trajectory collapses onto the unstable limit cycle. 

Notice that in this case the density of a homogeneous CA follows a limit cycle, oscillating between zero and one. 

\section{Numerical results for the majority rule}\label{sec:majority}

\begin{figure}[t]
    \centering
    \begin{tabular}{ccc}
     \raisebox{.5cm}{\includegraphics[width=0.2\linewidth]{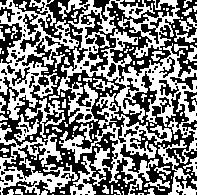}} &
     \includegraphics[width=0.25\linewidth]{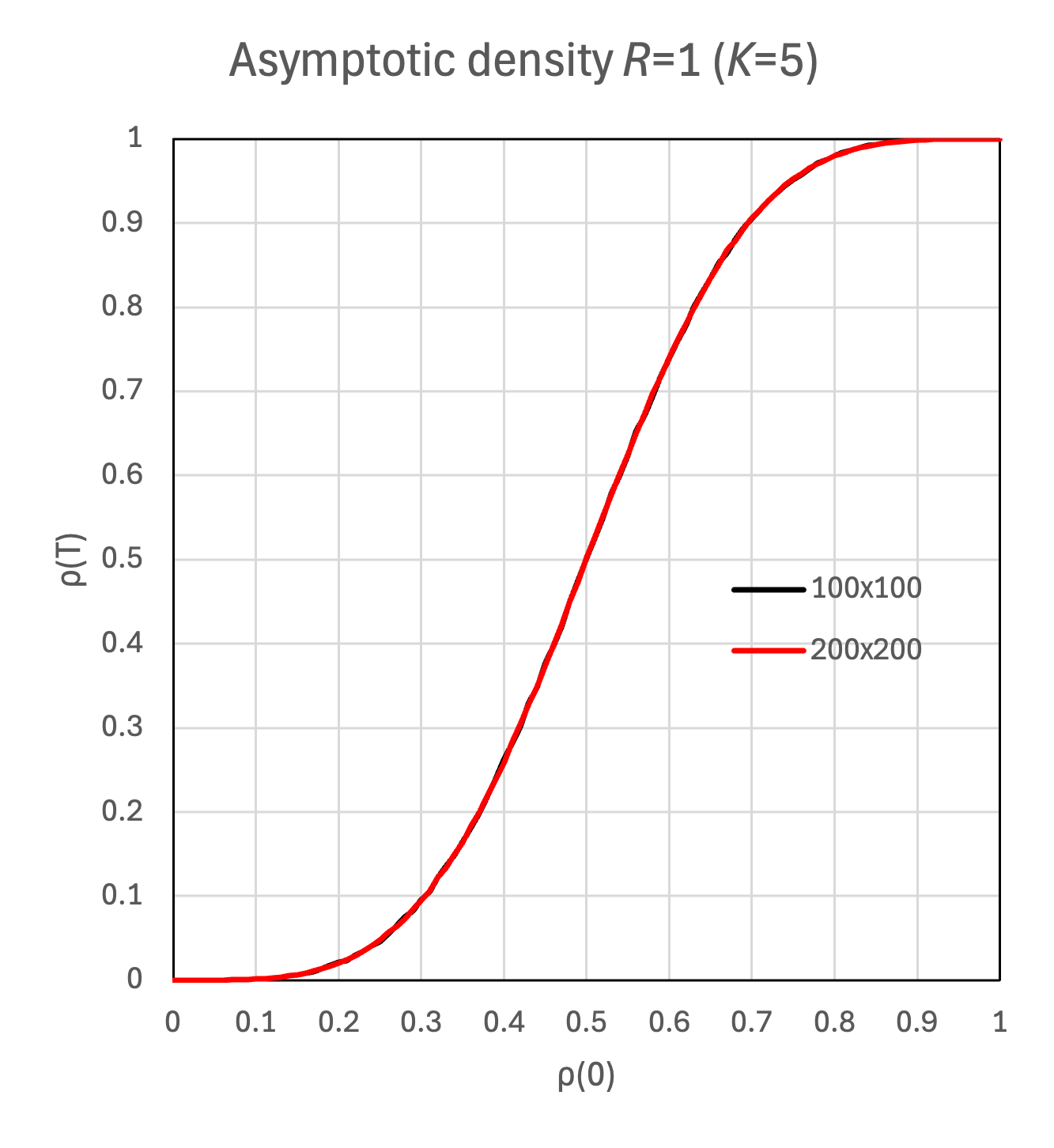} & 
    \includegraphics[width=0.25\linewidth]{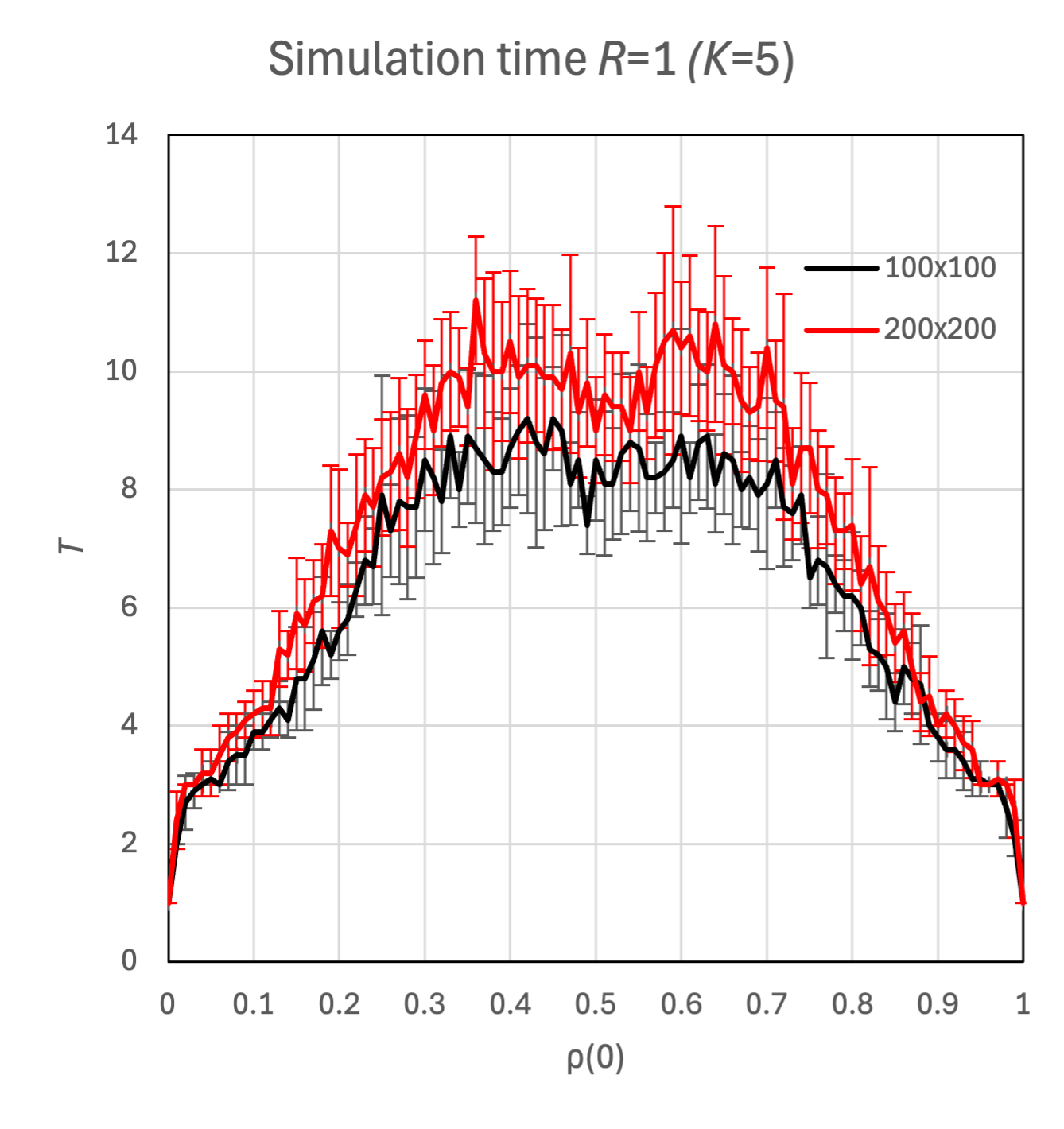}\\
    \raisebox{.5cm}{\includegraphics[width=0.2\linewidth]{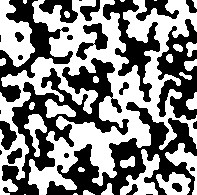}} &
    \includegraphics[width=0.25\linewidth]{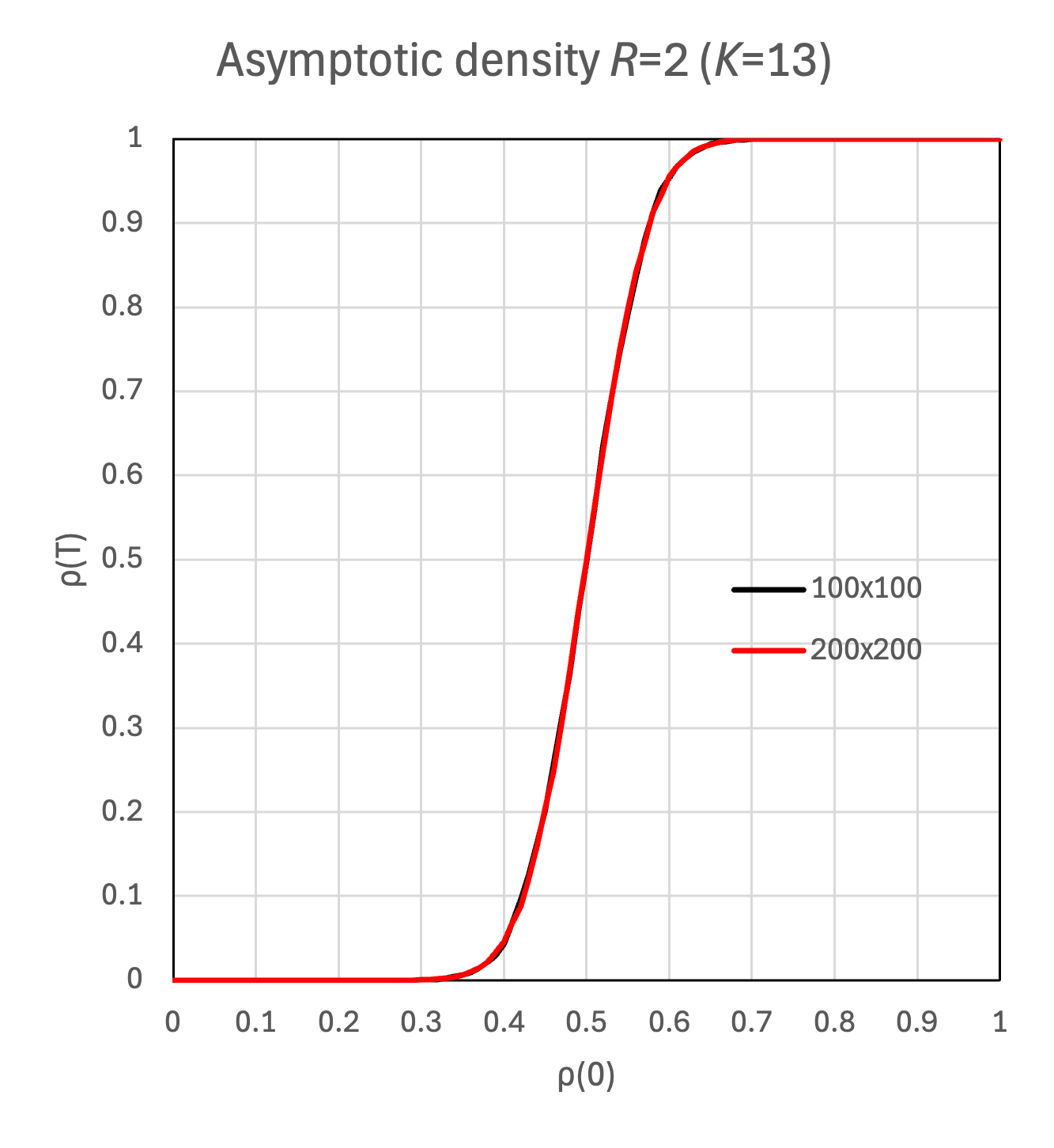} & 
    \includegraphics[width=0.25\linewidth]{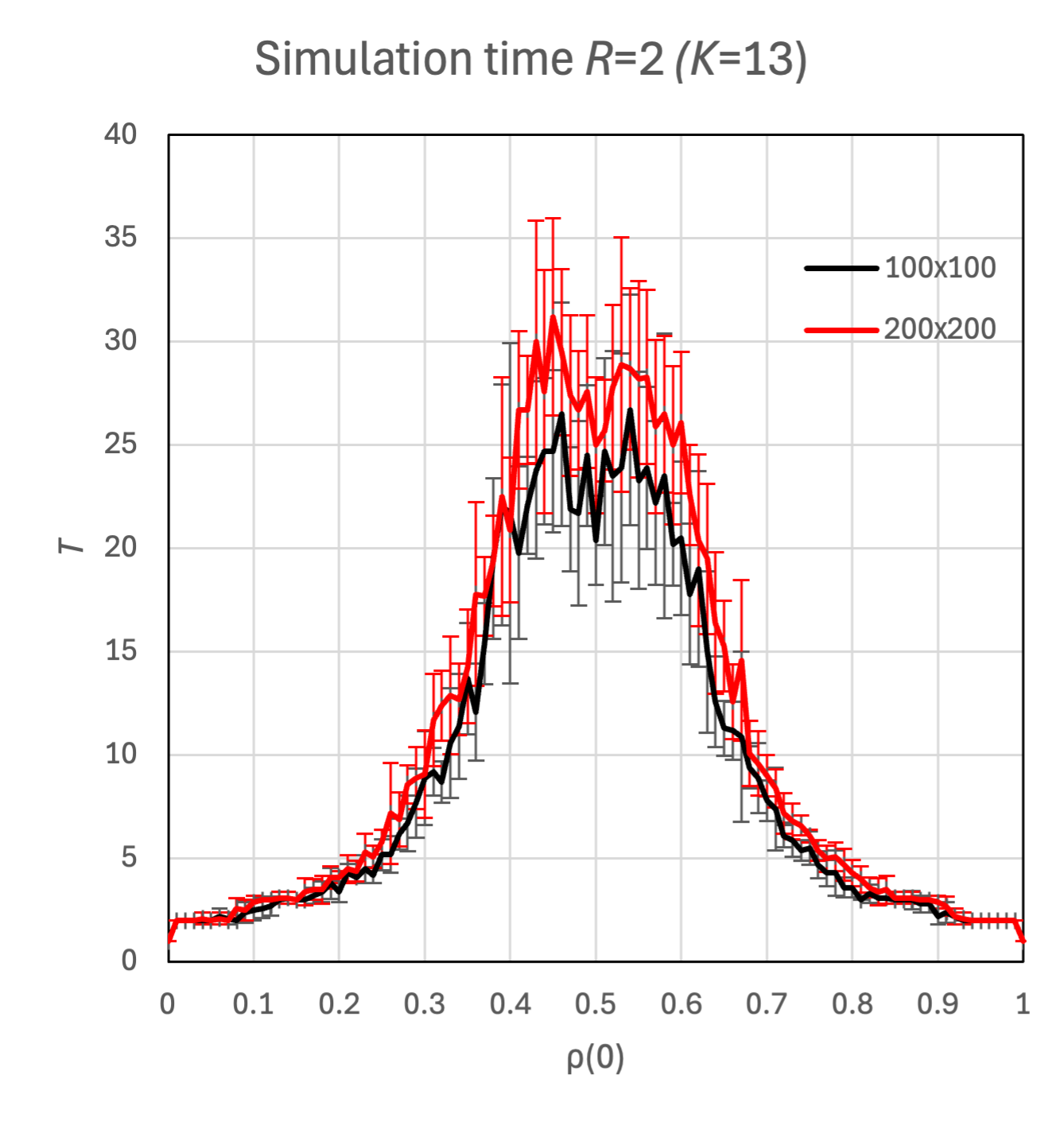}\\
    \raisebox{.5cm}{\includegraphics[width=0.2\linewidth]{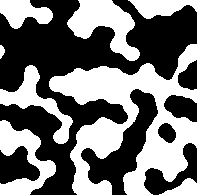}} &
    \includegraphics[width=0.25\linewidth]{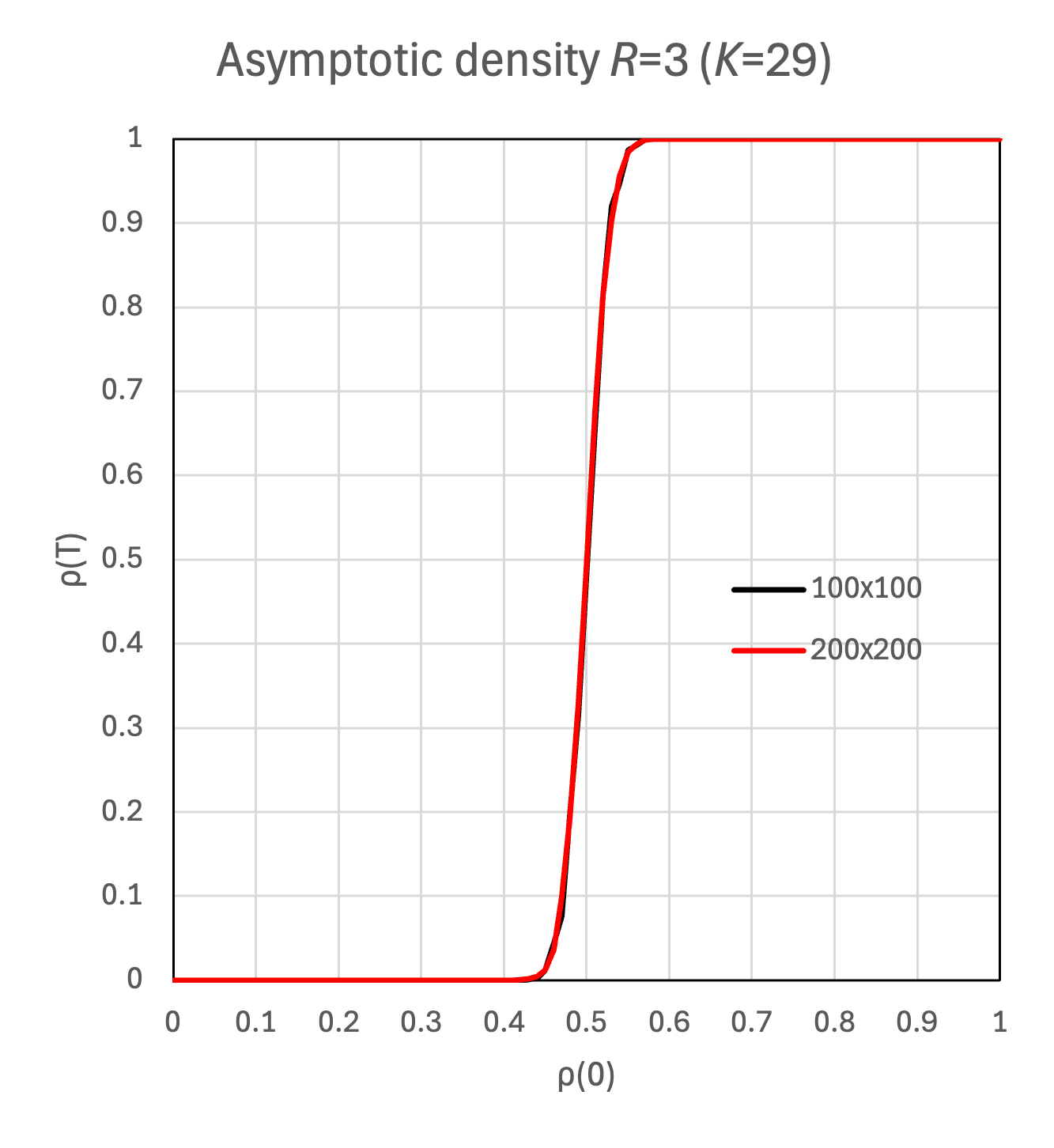} & 
    \includegraphics[width=0.25\linewidth]{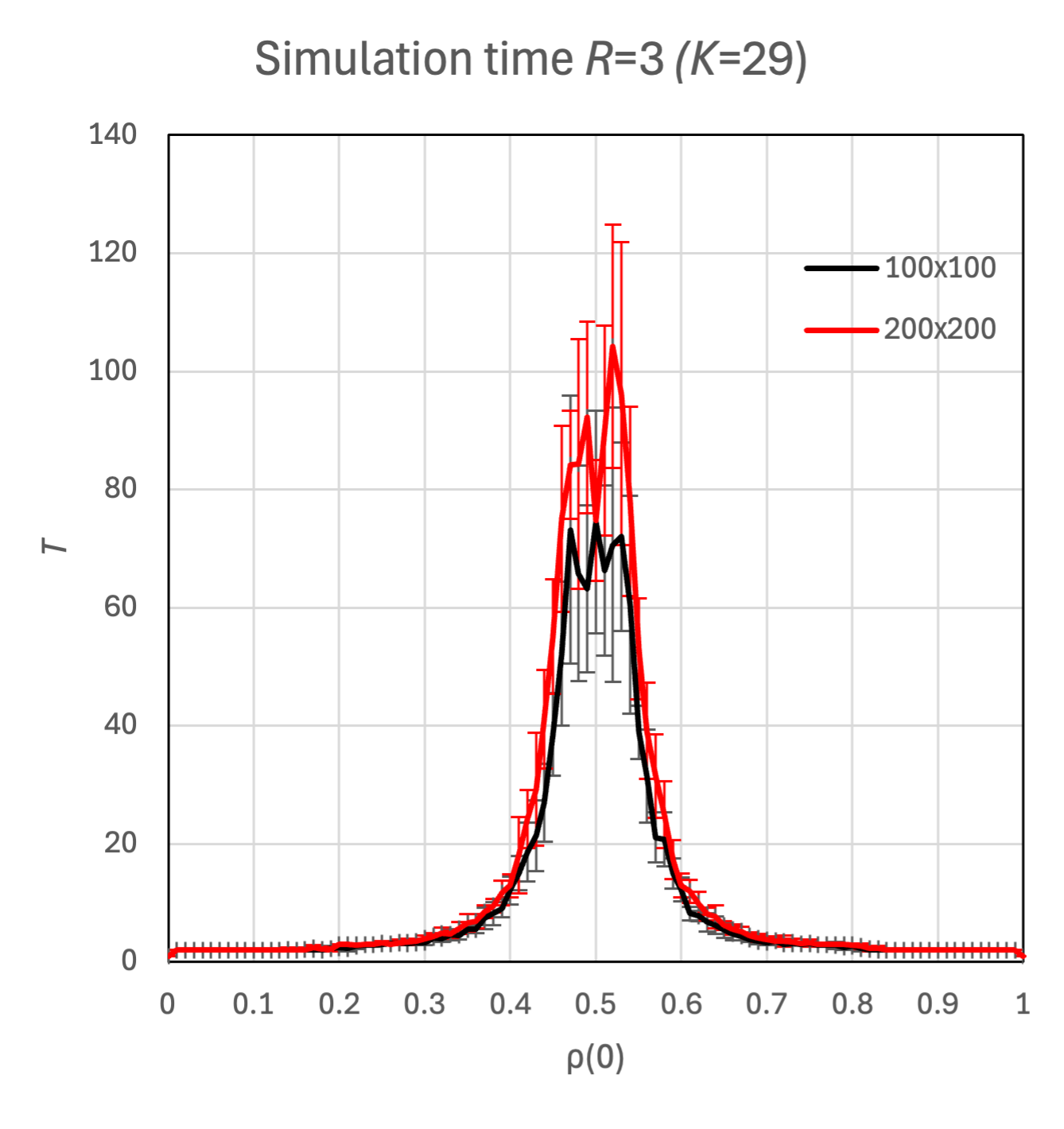}
    \end{tabular}
    \caption{Serial majority model for various values of $R$. Left column: the asymptotic configuration for a lattice $100\times 100$ cells and $\rho_0=0.5$. Center and right columns: average over 10 runs for a $100\times 100$  and $200\times 200$ lattices. Center column: the asymptotic density $\rho(T)$ vs initial density $\rho(0)$ (the values for the different lattice sizes are overlapped). Right column: the relaxation time $T$ versus the initial density $\rho(0)$. From top to bottom: $R=1$ ($K=5$); $R=2$ ($K=13$); $R=3$ ($K=29$).}
    \label{fig:R}
\end{figure}

Some simulations of the majority model of Eq.~\eqref{eq:majority} for a lattice of $100\times 100$ lattice sites are reported in Fig.~\ref{fig:R}. The results are essentially the same for the serial and parallel update, the only difference is that for parallel update one can have oscillating checkerboard patterns for certain values of $R$. For instance, for $R=1$, a cell in state zero is surrounded by four neighbors in state one, and a cell in state one by four neighbors in state zero, so they both flips their state. However, the appearance of such a pattern starting from a random configuration is quite rare. 

As shown in Fig.~\ref{fig:R}, the slope of the density curve increases with $R$, while it does not depend on the lattice size. One can safely assume that the transition is first-order for large values of $R$, the simulation time diverges at the transition in the same limit. 

As we can see from Fig.~\ref{fig:R}, left column, for $\rho_0=0.5$ the voter model almost always ends in a coexistence configuration, with clusters that evolve until the curvature radius $r$ of the boundary is above a certain threshold $r_c(R)$. 

\begin{figure}[t]
    \centering
    \begin{tabular}{cc} 
    \includegraphics[width=0.45\linewidth]{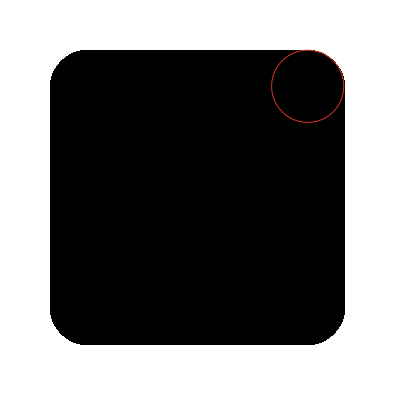} & 
 \includegraphics[width=0.5\linewidth]{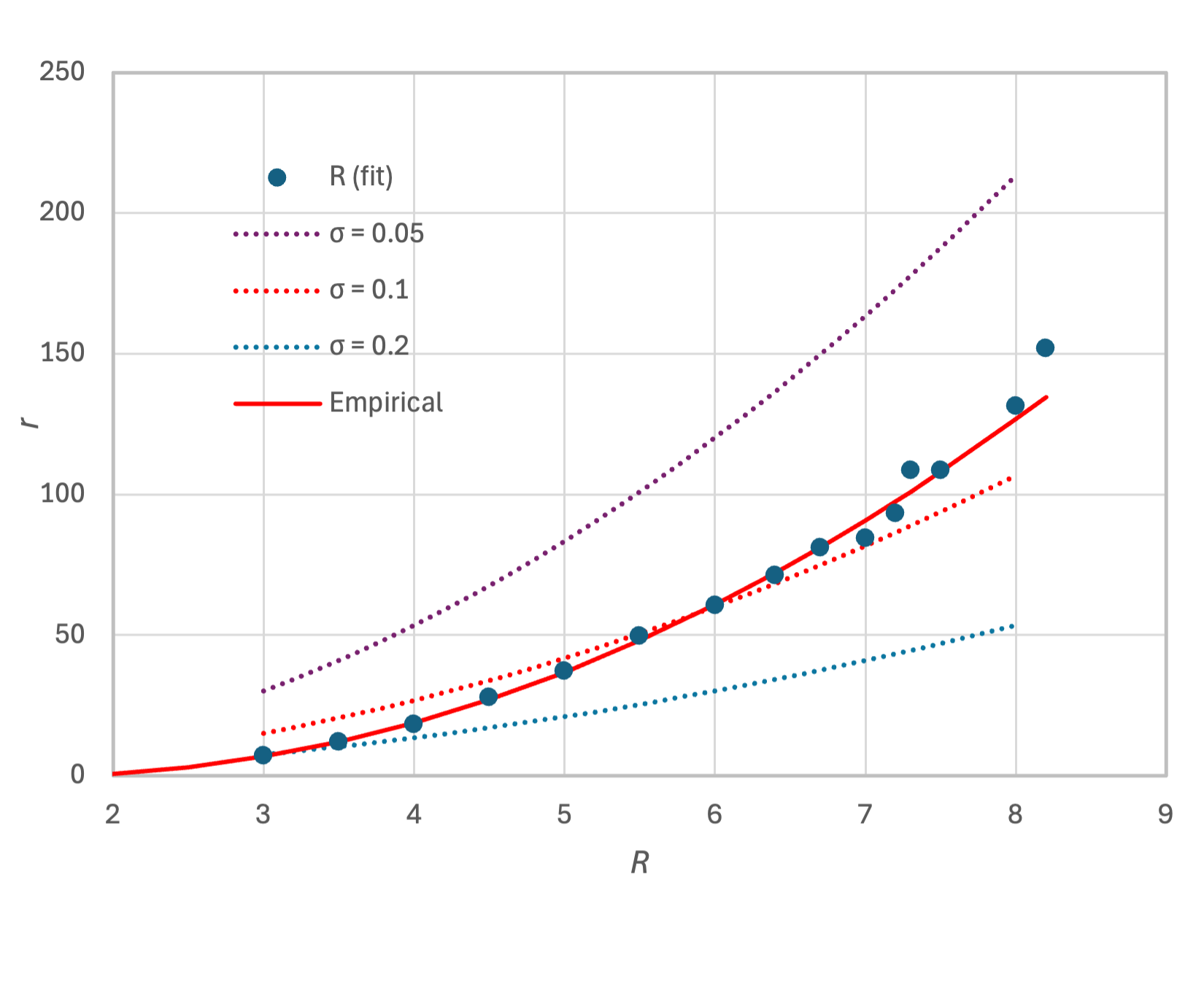}
 \end{tabular}
    \caption{Left: The procedure to numerical fit the curvature radius $r$ for a given interacting radius $R$. Right: the numerical fit of $r$ and the approximations for some values of $R$. Circles: The curvature radius $r$  numerically fitted for various interacting radius $R$. Red continuous line: the empirical fit $r = 3(R-1.5)^2$. Dotted curves: analytical approximations for different values of the parameter $\sigma$.}
    \label{fig:Rfit}
\end{figure}

We can experimentally measure the curvature radius $r$ for a given interacting radius $R$ by initializing the configuration with a large square of ones, surrounded by zeros. The system evolves until the corners are rounded above the critical value $r_c(R)$. At this point we can search for the largest radius, tangent to the horizontal and vertical boundaries of the one-state region and not including any zero-state cells, see Fig.~\ref{fig:Rfit}-left.

The results are reported in Fig.~\ref{fig:Rfit}-right, with some approximations illustrated in the following. One can notice that, although there is an evident trend (roughly $r = 3(R-1.5)^2$, red continuous line in Fig.~\ref{fig:Rfit}-right), fluctuations are always present also for large values of $R$.

\begin{figure}[t]
    \centering
    \begin{tabular}{cc}
    \includegraphics[width=0.4\linewidth]{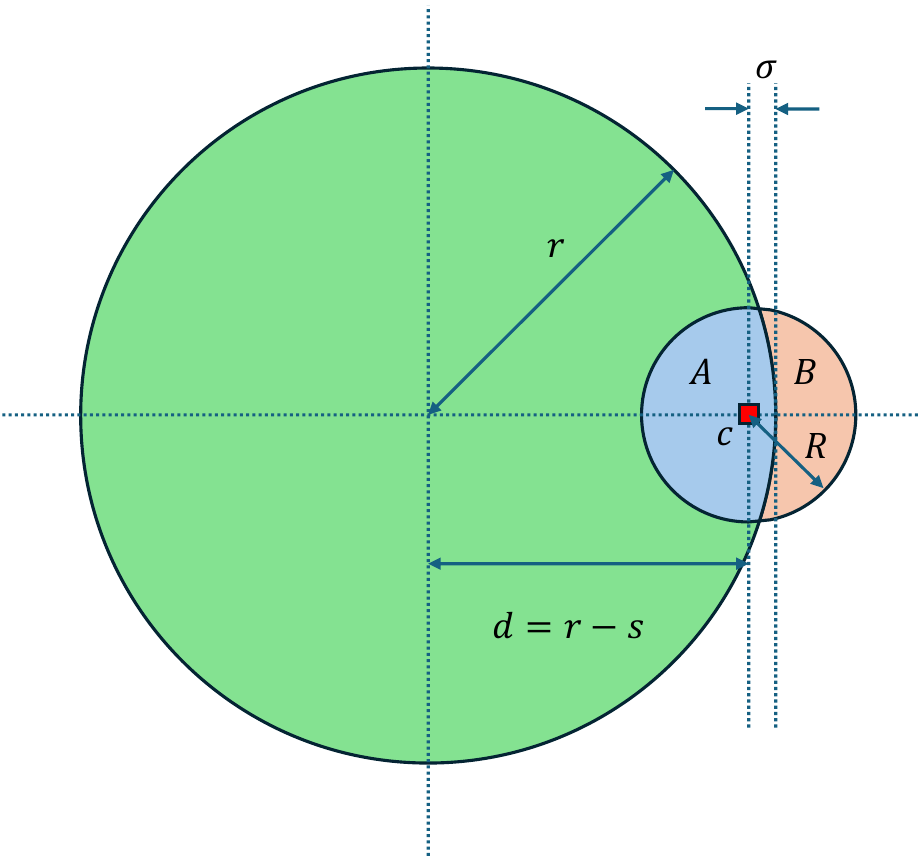} &
    
    \includegraphics[width=0.5\linewidth]{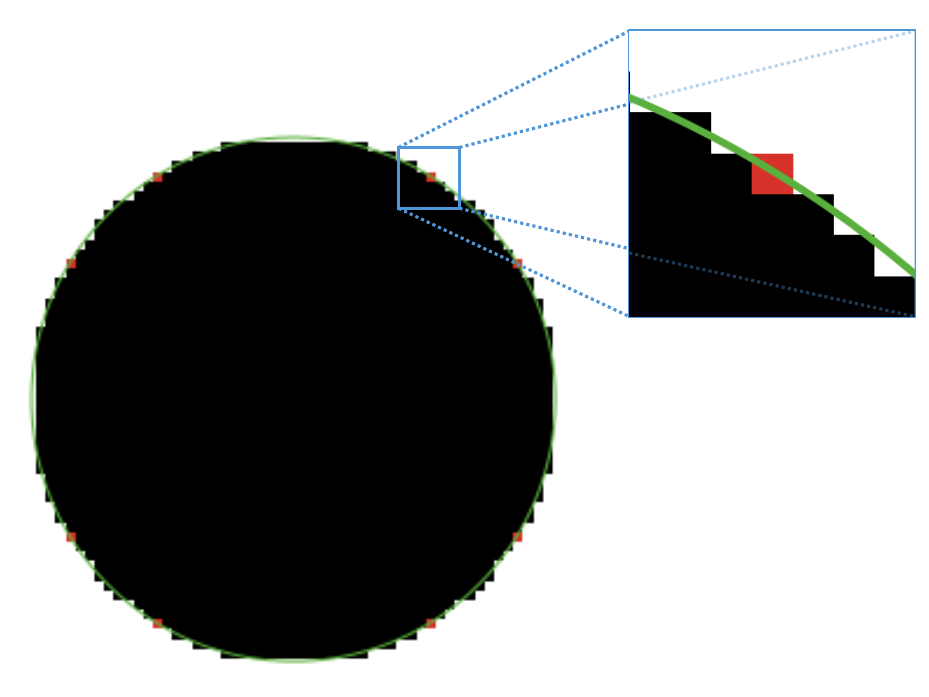} 
    \end{tabular}

    \caption{Right: The point $c$ keeps the state of other sites in the green radius if the area $A$ is larger than the area $B$. Left: The variation of the number of grid points in a circle of radius $r=12.9$ after increasing the interacting radius  $R=3.8$ by the minimal amount $\sigma=0.1$ }
    \label{fig:scheme}
\end{figure}

The relation between the curvature radius $r$ and the interacting radius $R$ can be derived in the continuous approximation as shown in Fig.~\ref{fig:scheme}-left.

A given cell $c$ belonging to the border of a cluster keeps the same state $s$ if in its neighborhood the majority of  cells are in the same state. If we assume that  the cluster has locally a curvature radius $r$, and calling $\sigma$ the distance between cell $c$ and the first cell with opposite state in the direction perpendicular to the boundary, the previous statement corresponds to say that the area $A$ of the intersection between the circle centered in $c$ with radius $R$ and the circle of radius $r$, in the direction perpendicular to the surface and at a distance $d=r-\sigma$ is greater than $B=\pi R^2 - A$. 

The intersection $I$ between the two circles is 
\eq{
  \begin{split}
    I(r) &= r^2 \cos^{-1} \left(\frac{r^2 + d^2 - R^2}{2dr}\right) + R^2 \cos^{-1} \left(\frac{R^2 + d^2 
    - r^2}{2dr}\right) 
    \\
    &- \frac{1}{2}\sqrt{(r+d-R)(R+d-r)(R+r-d)(R+r+d)},
  \end{split}
}
so the above condition becomes, at the threshold, 
\eq{
    I(r_c)=\frac{\pi R^2}{2} 
}

The problem is that the correct value of $\sigma$ depends on $r$, since its location on the circle is not easily determined. This problem is related to the Gauss circle one, i.e., $\sigma$ is the minimal increment of $r$ such that the number of grid points included in the circle changes, see Fig:~\ref{fig:scheme}-right. 

Indeed, one can see in Fig.~\ref{fig:Rfit} that no analytical approximation fits all numerical points for all values of $R$: for  $R\le 4$ a value $\sigma=0.2$ is more plausible, for $4<R\le 7$ a value $\sigma=0.1$ is more reasonable, and then the most adequate distance probably diminishes.

\section{Numerical results for the frustrated majority rule}\label{sec:frustrated}

The frustrated majority rule of Eq.~\eqref{eq:frustrated} is interesting because it also behaves quite differently from what expected by the mean-field approximation, Fig.~\ref{fig:frustratedmf}.

\begin{figure}[t]
    \centering
    \begin{tabular}{cc}
    \includegraphics[width=0.33\linewidth]{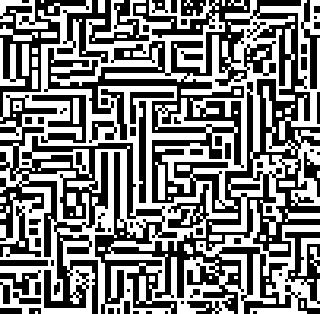} & 
    \includegraphics[width=0.33\linewidth]{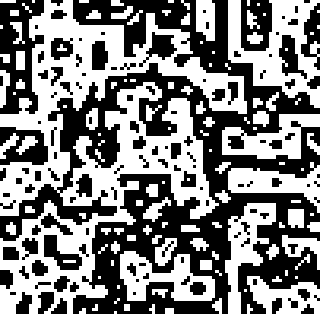} \\
    \includegraphics[width=0.33\linewidth]{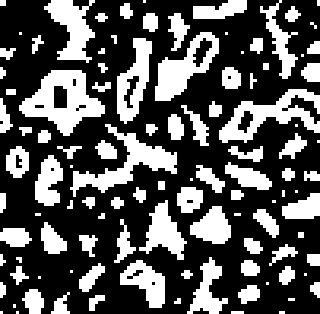} & 
    \includegraphics[width=0.33\linewidth]{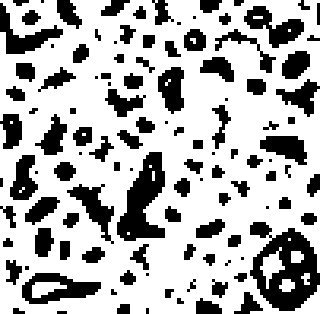} 
    \end{tabular}
    
    \caption{Typical patters of the frustrated majority problem of Eq.~\eqref{eq:frustrated}for a $100\times100$ lattice. Top-left: $R=1$, top-right: $R-2$; bottom: $R=3$ starting with (left) $\rho_0=0.1$ and (right) $\rho_0=0.9$.
    }
    \label{fig:frustratedR}
\end{figure}

The  simulations show time-changing patterns for the configuration $@s$, with a well defined density, see Fig.~\ref{fig:frustratedR}. Only for very large values of $R$ and very low or very high values of the initial density $\rho_0$ one can see the oscillation between all-zero and all-one states, but this is due to the difficulty in forming a "seed" of different density with respect to average one, related to the finiteness of the lattice. If we provide a localized perturbation in the density, the system evolves towards a mixed state. 

\begin{figure}[t] 
\centering
\includegraphics[width=0.9\linewidth]{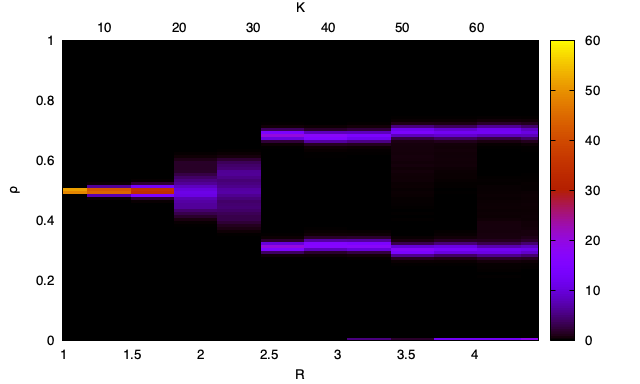}
    \caption{The bifurcation diagram of the probability distribution of the asymptotic density $\rho$ vs. $\rho_0$ for various values of the interacting radius $R$. Lattocie size $200\times200$ cells, 32 samples with $\rho_0$ varing from $1/33$ to $32/33$, averaged over $10^3$ time steps after a transient of $4\cdot10^3$ steps. }
    \label{fig:bifurcation}
\end{figure}

What is particular, above $R=2.5$ the asymptotic density depends (inversely) from the initial one, $\rho_0$. Indeed, the probability distribution of the asymptotic density $\rho$ as a function of $\rho_0$ for different values of $R$ shows a bifurcation diagram, as reported in Fig.~\ref{fig:bifurcation}. 

We do not have any theoretical interpretation for this behavior. 

\section{Conclusions}\label{sec:conclusion}
We have investigated Boolean, totalistic cellular automata with a majority rule and extended interaction range $R$. 

The mean-field approximation of this model, which is analogous to the voter one, gives for the asymptotic density only two extreme values, zero and one (absorbing states), depending on the initial density $\rho_0$.

For initial density equal to $0.5$, which should constitute an unstable separation between the basins of the two absorbing state, we have shown that  the dynamics is dominated by coarsening. The final patterns are constituted by clusters with a definite curvature radius $r$. We have elaborated a continuous approximation for the relationship between $R$ and $r$, but we have shown that the free parameter of this approximation cannot take a unique value for all values of $R$. 

For the frustrated majority model, which have an absorbing limit cycle, 
the mean-field approximation gives chaotic oscillations or the mentioned limit cycle. Instead, we observed active patterns, with a well-defined and time-stable density. 

Above a certain critical value of the interacting radius we observed the appearance of a bifurcation of the asymptotic density as a function of the initial one. When $\rho_0$ is smaller than $1/2$ the evolution of the system gives origin to patterns with asymptotic density larger than $1/2$, and vice versa.  

We think that these two problems deserve further studies. 
\printbibliography

\end{document}